\documentclass[]{spie}  

\usepackage{amsmath,amsfonts,amssymb}
\usepackage{graphicx}
\usepackage[colorlinks=true, allcolors=blue]{hyperref}

\title{Athena Wide Field Imager Key Science Drivers}

\author[a]{Arne Rau}
\author[a]{Kirpal Nandra}
\author[b]{James Aird}
\author[c]{Andrea Comastri}
\author[d]{Thomas Dauser}
\author[a]{Andrea Merloni}
\author[e]{Gabriel W. Pratt}
\author[f]{Thomas H. Reiprich}
\author[b]{Andy C. Fabian}
\author[a]{Antonis Georgakakis}
\author[g]{Manuel G\"udel}
\author[h]{Agatha R\'o\.za\'nska}
\author[a]{Jeremy S. Sanders}
\author[i]{Manami Sasaki}
\author[j]{Simon Vaughan}
\author[d]{J\"orn Wilms}
\author[a]{Norbert Meidinger}

\affil[a]{Max Planck Institute for extraterrestrial Physics, Giessenbachstr. 1, Garching, Germany}
\affil[b]{Institute of Astronomy, University of Cambridge, Madingley Road, Cambridge, CB3 0HA, UK}
\affil[c]{INAF Osservatorio Astronomico di Bologna, via Ranzani 1, I-40127, Bologna, Italy}
\affil[d]{Remeis Observatory \& ECAP, Universit\"at Erlangen-N\"urnberg, Sternwartstr. 7, 96049, Bamberg, Germany}
\affil[e]{Laboratoire AIM, IRFU/Service d'Astrophysique - CEA/DRF - CNRS - Université Paris Diderot, at. 709, CEA-Saclay, 91191, Gif-sur-Yvette Cedex, France}
\affil[f]{Argelander Institute for Astronomy, Bonn University, Auf dem H\"ugel 71, 53121 Bonn, Germany}
\affil[g]{Department of Astronomy, University of Vienna, Vienna, Austria}
\affil[h]{N. Copernicus Astronomical Center, Bartycka 18, 00-716, Warsaw, Poland}
\affil[i]{Institut f\"ur Astronomie und Astrophysik, Universit\"at T\"ubingen, Sand 1, 72076, T\"ubingen, Germany}
\affil[j]{X-ray \& Observational Astronomy Group, Department of Physics and Astronomy, University of Leicester, Leicester LE1 7RH, UK}

\authorinfo{Further author information: (Send correspondence to A.Rau: arau@mpe.mpg.de)}

\begin{document} 
\maketitle

\begin{abstract}
The Wide Field Imager (WFI) is one of two instruments for the Advanced Telescope for High-ENergy Astrophysics (Athena). In this paper we summarise three of the many key science objectives for the WFI -- the formation and growth of supermassive black holes, non-gravitational heating in clusters of galaxies, and spin measurements of stellar mass black holes -- and describe their translation into the science requirements and ultimately instrument requirements. The WFI will be designed to provide excellent point source sensitivity and grasp for performing wide area surveys, surface brightness sensitivity, survey power, and absolute temperature and density calibration for in-depth studies of the outskirts of nearby clusters of galaxies and very good high-count rate capability,  throughput, and low pile-up, paired with very good spectral resolution, for detailed explorations of bright Galactic compact objects.  
\end{abstract}


\section{INTRODUCTION}
\label{sec:intro}  

The Advanced Telescope for High-ENergy Astrophysics ({\it Athena}) is ESA's next large space observatory for the exploration of the X-ray sky. The mission, anticipated to be launched in 2028, will primarily be designed to address two fundamental questions of astrophysics, namely "How do black holes grow and shape the Universe?" and "How does ordinary matter assemble into the large scale structures that we see today?" [\citenum{Nandra:2013aa}]. As an observatory, {\it Athena} will provide a unique contribution to astrophysics in the 2030s by exploring high-energy phenomena in all astrophysical contexts, including those yet to be discovered (see also [\citenum{Nandra:2016aa}]). 

\begin{figure} [ht]
\begin{center}
\begin{tabular}{cc} 
\includegraphics[height=8.2cm]{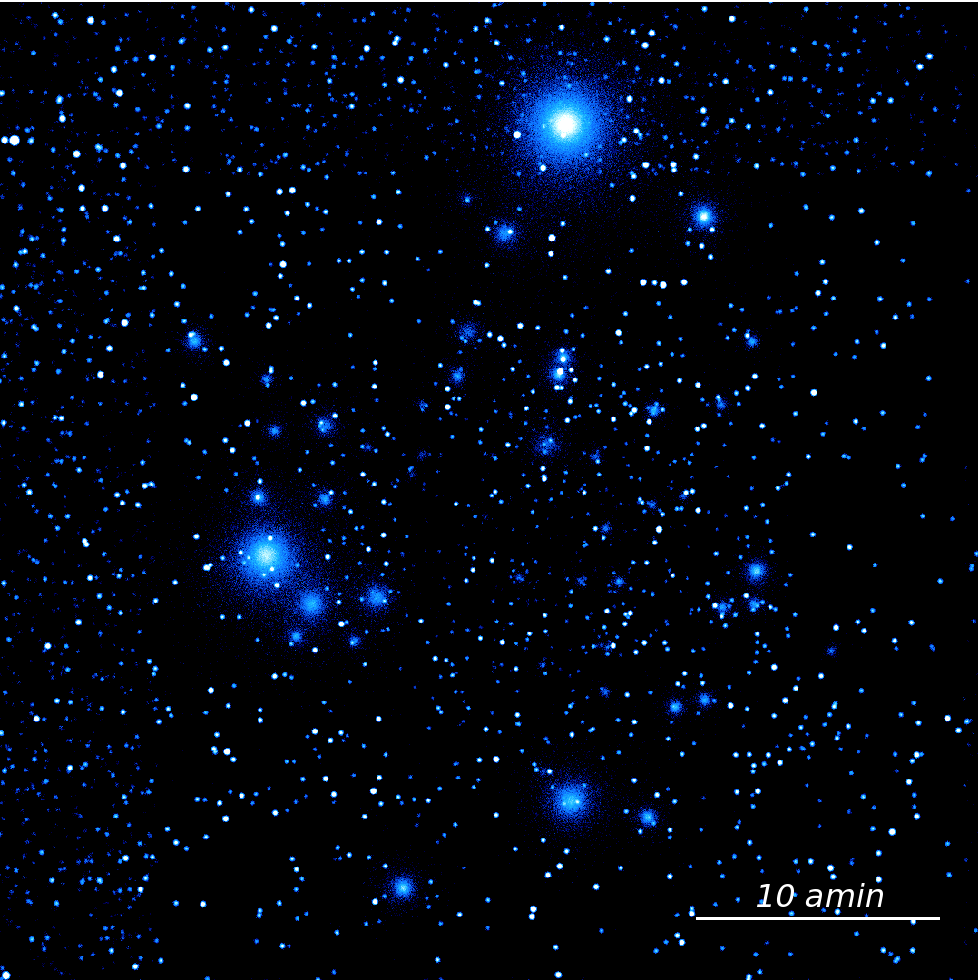}
\includegraphics[height=8.2cm]{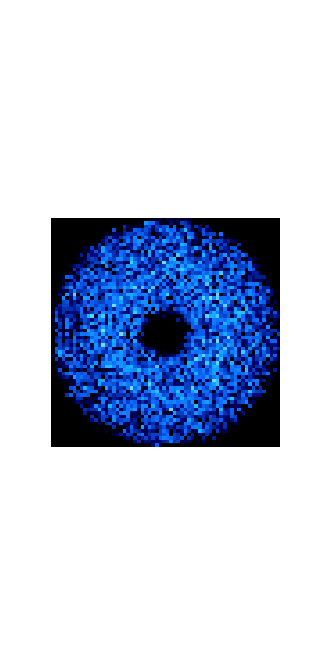}
\end{tabular}
\end{center}
\caption[example]{{\it Left:} 100\,ks SIXTE (http://www.sternwarte.uni-erlangen.de/$\sim$sixte) simulation of an observation with the WFI large field of view detector. The image shows the composition of the source population of the Chandra Deep Field South observed with {\it Chandra} and a set of extended objects representing clusters of galaxies. {\it Right:} Simulation of an observation of a bright point source with the WFI fast detector. The defocusing leads to a donut-like photon distribution that mimics the geometry of the {\it Athena} mirror assembly. Note, that the image is enlarged by $\sim\times6$ and does not represent the true scale compared to the large detector.
 \label{fig:fov} }
\end{figure}

The performance required to achieve the science goals will be realised through the combination of state-of-the-art Silicon Pore Optics (SPO [\citenum{Bavdaz:2016aa}]) mirror technology and two complimentary cutting-edge scientific instruments. Unprecedented very-high resolution X-ray spectroscopy ($\Delta E\le2.5$\,eV at 7\,keV)  will be provided by the X-ray Integral Field Unit (X-IFU, see [\citenum{Barret:2016aa}] for more details) over a $5^{\prime}$ diameter Field of View (FoV). The Wide Field Imager (WFI, see [\citenum{Meidinger:2016aa}] for more details) will offer unrivaled survey power with X-ray imaging at lower spectral resolution ($\Delta E<\le170$\,eV at 7\,keV) but with very good spatial resolution ($\le5^{\prime\prime}$ PSF on-axis) over a larger FoV of $40^{\prime}\times40^{\prime}$. In addition, a WFI dedicated observing mode will enable  excellent high-count rate capabilities for observations of bright point sources. See Fig.~\ref{fig:fov} for two simulation of typical observations with the large and fast detector.

Here we summarise the key science drivers for the WFI and the derived requirements on the instrument performance parameters.

\section{The Wide Field Imager in one paragraph}
\label{sec:wfi_instrument}

Instead of starting directly with the {\it Athena} science goals and subsequently motivating the WFI parameters, we begin here with one-paragraph summary of the main instrument characteristics. This provides the reader with a useful initial picture of the WFI and at the same time acts as a reminder of the tight link between the scientifically driven requirements and the technical feasibility, i.e. the technical possibilities set the boundaries on which the science requirements push. A detailed description of the instrument and its development status can be found in [\citenum{Meidinger:2016aa}].

The WFI consists of two independently operated detectors, a large detector for wide field applications and a separate small detector optimised for high-time resolution observations of bright point sources. Both detectors are based on the same DEPFET\footnote{Depleted p-channel field effect transistor [\citenum{Kemmer:1987aa}]} active pixel sensor technology with a pixel size of $130\,\mu$m. The energy range of interest is $0.2-15$\,keV with a spectral resolution of $<170$\,eV at 7\,keV. The large detector consists of a $2\times2$ matrix of $512\times512$ pixels filling a  $40^{\prime}\times40^{\prime}$ FoV with a scale of $\sim2.24^{\prime\prime}$ pxl$^{-1}$. The small detector has $64\times64$ pixels and will be operated out-of-focus,  i.e. without imaging capability, to optimise the pile-up and throughput performance for point sources. It will be read out in two halves simultaneously with a frame time of $80\,\mu$s. Each of the two detectors will have its own external filter and calibration source in a common filter wheel.  At any given time, the {\it Athena} optics will illuminate only one of the two detectors (or the X-IFU) and the exchange will be performed by tilting the mirror assembly. 
 
\section{WFI science drivers}
\label{sec:science_drivers}

With the basic picture of the instrument presented in \S~\ref{sec:wfi_instrument}, we can now proceed with highlighting some of the main science drivers for the WFI. As a complete census of the WFI science objectives is clearly beyond the scope of this paper, we focus in particular on three topics here, the formation and growth of supermassive black holes  (\S~\ref{subsec:agn}), the hot gas in early galaxy groups and in galaxy clusters (\S~\ref{subsec:clusters}), and spin measurements of Galactic black hole binaries (\S~\ref{subsec:spin}).

\subsection{The Formation and Growth of Supermassive Black Holes}
\label{subsec:agn}

Supermassive black holes (SMBHs) with masses of several million to billion times that of the Sun exist in the centres of all large galaxies. Their formation and growth mechanisms are unsolved puzzles now and will very likely remain challenges until the time of {\it Athena} [\citenum{Aird:2013aa}]. The main obstacle here is the lack of access to suitable probes of the infant Universe, i.e. a large sample of active galactic nuclei (AGN) at a redshift of $z>6$. Only a few tens of the most luminous ($L_X\sim10^{47}$\, erg/s) and most heavy-weight sources are known from optical surveys (e.g., Sloan, [\citenum{Fan:2003aa}]) beyond that redshift. These objects, however, have exceptional characteristics and thus are not representative of the typical AGN population at these early epoch. They must have increased their masses by many orders of magnitude in a very short time, erasing any potential tracers of the initial formation conditions. {\it Athena}/WFI observations will instead uncover the X-ray emission of objects at the fainter end of the AGN luminosity function ($10^{43}<L_X<10^{44}$\,erg/s at $z\sim6-7$). This population is much more numerous than the exceptionally bright, already known sources and overall dominates the accretion power output in the early Universe. 

\begin{figure} [ht]
\begin{center}
\begin{tabular}{cc} 
\includegraphics[height=7.5cm]{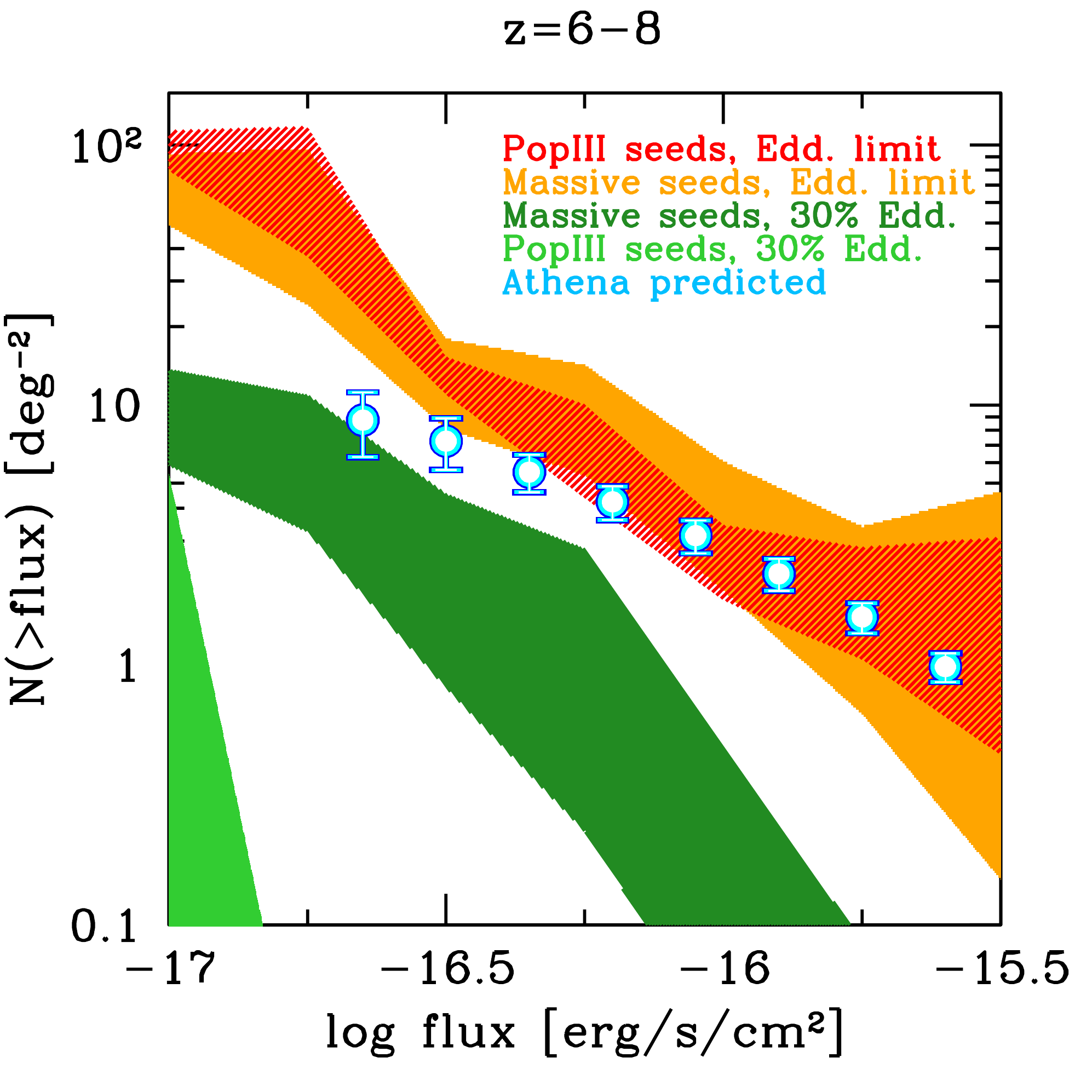}
\includegraphics[height=7.5cm]{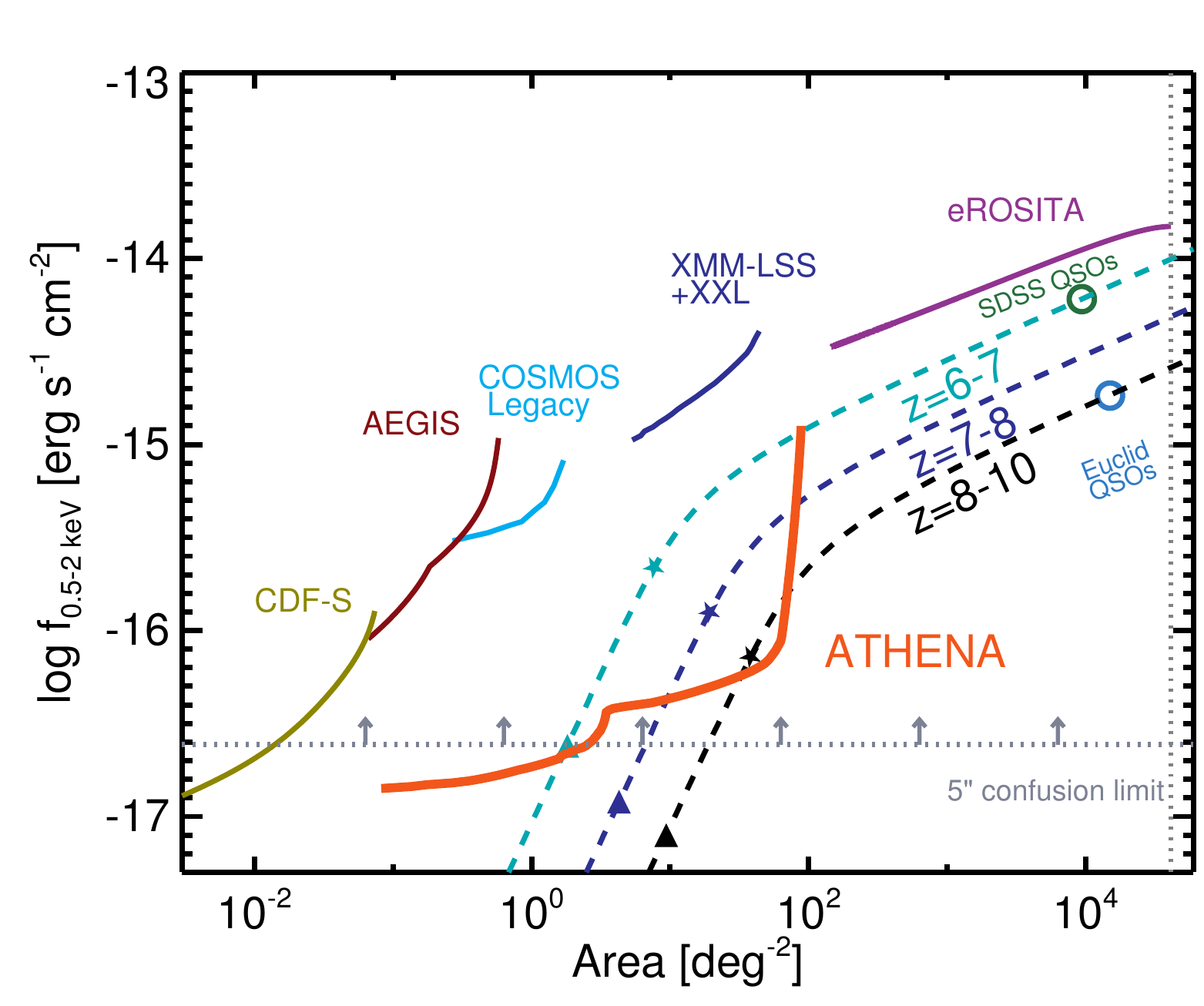}
\end{tabular}
\end{center}
\caption[example] {{\it Left:} Forecast for the $z\sim6-8$ AGN number counts detectable with the WFI (blue circles) compared to theoretical predictions for different black hole seed formation and growth models (filled regions). For clarity, only the extreme ends of the seed masses (PoP III \& primordial gas cloud collapse) are depicted here.The predicted WFI measurements are based on extrapolations from lower redshift data. Figure updated from [\citenum{Aird:2013aa}]. {\it Right:} 0.5-2\,keV flux limits as function of area for a potential realisation of the multi-tiered WFI survey compared with current surveys. Dashed lines mark the parameter spaces to find $>10$ sources in a given redshift interval. The orange line shows the sensitivity achieved in $16\times450$\,ks and $230\times80$\,ks. Figure updated from [\citenum{Aird:2013aa}].
\label{fig:bh_seeds} }
\end{figure}

Measuring the X-ray luminosity function with the detection of several hundred AGN at $z>6$ will allow to constrain the seeds and processes that led to the early growth of SMBHs. Three possible formation mechanisms have been proposed (see also [\citenum{Volonteri:2012aa}]): i) The deaths of the first generation of massive stars (Population III) could have left $\sim100$\,M$_{\odot}$ remnants, which would have to grow very rapidly, e.g., through merging with other remnants or by accretion, to reach the observed $10^9$\,M$_{\odot}$ by $z\sim6$ [\citenum{Madau:2001aa}]. ii) The collapse of massive star clusters in the centres of the first galaxies may have produced $\sim1000$\,M$_{\odot}$ seed black holes [\citenum{Devecchi:2009aa}]. iii) Even more massive seeds ($10^5$\,M$_{\odot}$) with weaker demands on the growing rate can have formed in the direct collapse of primordial gas clouds [\citenum{Begelman:2006aa}]. Depending on the formation epoch and accretion history each of the above listed formation models predicts different AGN number counts per unit area as function of luminosity at any given redshift, i.e. age of the Universe. The distribution of AGN to be detected with the WFI will be compared with these theoretically motivated forecasts to identify the best fitting scenario (Fig.~\ref{fig:bh_seeds}, left). For example, a low number density of $z\sim6-8$ sources would imply a low growth rate and thus be incompatible with Population III seeds accreting near the Eddington limit.

The necessary sample ($>400$ AGN at $z>6$ and $>20$ AGN at $z>8$) is expected to be uncovered in the deep areas of a multi-tiered WFI survey (see also Fig.~\ref{fig:bh_seeds}, right). This survey will address many different {\it Athena} science goals by combining very sensitive observations over small areas with shallower observations over a larger part of the sky \footnote{A possible realisation would be the combination of $230\times80$\,ks, $9\times450$\,ks, $3\times700$\,ks, and $4\times1$\,Ms observations. The {\it Athena}/WFI survey speed will be $100\times$ faster than currently possible with e.g., {\it Chandra} or {\it XMM-Newton}}. The high-redshift AGN search places in particular a requirement on the grasp (the product of effective area and field of view). The targets are faint and rare and thus a large sky area needs to be observed to great depth. Distinguishing the faint AGN from fore- and background sources requires an excellent point source sensitivity over the entire field of view, which too depends on the effective area but also on the background (e.g., X-ray stray light) and the point spread function, i.e angular resolution. A precise knowledge of the astrometric uncertainty ($<1^{\prime\prime}$ at $3\,\sigma$) is needed to reliably associate multi-wavelength counterparts with the X-ray detections. The relation between the key requirements is sketched in Fig.~\ref{fig:bh_requirements}. The requirement values are summarised in Table~\ref{tab:requirements}.
 
\begin{figure} [ht]
\begin{center}
\begin{tabular}{cc} 
\includegraphics[height=4cm]{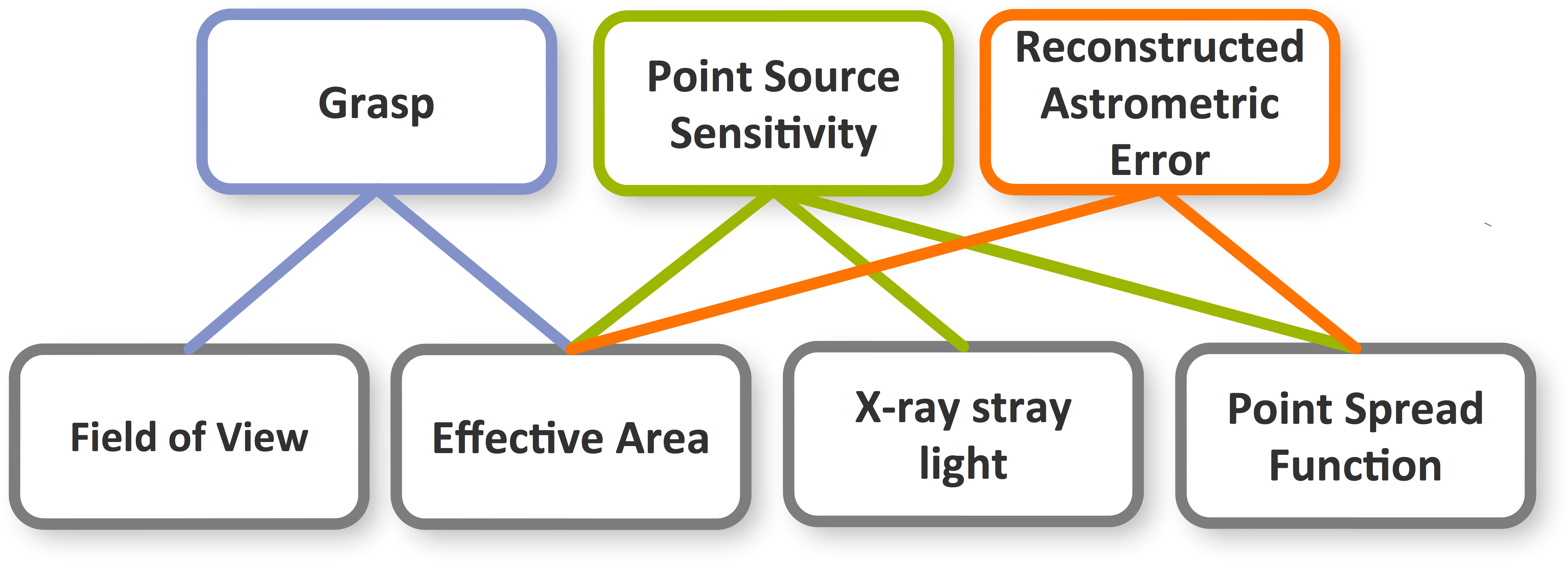}
\end{tabular}
\end{center}
\caption[example] {Relationships between the most driving requirements for the WFI high-redshift AGN science goal. The top row shows the mission independent performance requirements while the bottom row contains the derived mission/instrument specific core science requirements. Notes: point source sensitivity refers here to the sensitivity over a large section of the FoV; reconstructed astrometric error will be achieved by alignment with optical/near-IR catalogs. Detailed values are presented in Table~\ref{tab:requirements}.
\label{fig:bh_requirements} }
\end{figure}


\subsection{Non-gravitational Heating in Galaxy Clusters}
\label{subsec:clusters}

Located in the nodes of the Cosmic Web, clusters of galaxies are prime sites for investigating the formation and growth of the Universe's large-scale structure. As more than $>85$\,\% of their baryonic mass\footnote{which itself is only $11$\,\% of the total cluster mass, the rest being in the form of dark matter ($\sim90$\,\%)} is in the diffuse, hot X-ray emitting Intra-Cluster Medium (ICM), galaxy clusters are natural targets for the {\it Athena} mission. One of the main science objectives for the WFI  is to probe the hierarchical gravitational assembly process and to determine the physical processes that dominate the injection of non-gravitational energy (i.e., energy released not as a result of the gravitational collapse of the gas into the dark matter potential) into the ICM as function of cosmic epoch [\citenum{Ettori:2013aa}]. These non-gravitational processes include, e.g., the feedback produced by the powerful outflows of supernova, or heating by jets from the central AGN.

To investigate the gravitational heating of the gas during its collapse into the dark matter potential, and to  differentiate between the sources and models for non-gravitational energy input, studies up to and beyond the virial radius of $R_{200}$ in local clusters and out to R$_{500}$\footnote{R$_{200}$ or R$_{500}$ are the radii of a sphere centred on the galaxy cluster in which the average matter density is $200\times$ or $500\times$ the critical density of the Universe.} up to a redshift of $\sim2$ are required. The regions beyond $R_{500}$, which are commonly referred to as cluster outskirts, encompass nearly 80-90\,\% of the cluster volume. However, they are so far poorly explored as the current instrumentation limits their access to only a few of the brightest and nearby objects [e.g., \citenum{Pratt:2010aa,Eckert:2013aa}].

Entropy \footnote{defined as $K=kT/n^{2/3}_e$, with $T$ being the gas temperature and $n_e$ the number density of electrons.} is the key to understanding both gravitational heating of the gas and its modification by non-gravitational processes. 
Entropy is generated by shocks and compression of the gas during the hierarchical assembly process, but can be modified by other processes that change its physical characteristics. 
The radial distribution of entropy that is generated in a pure gravitational collapse can be predicted straight forwardly (e.g., [\citenum{Voit:2005aa}]). In the inner regions, deviations from this distribution originate in non-gravitational processes such as AGN feedback and gas cooling. In the outskirts, deviations originate in gas clumping as it is accreted into the cluster potential. The evolution of the entropy profile and its distribution give critical clues to its origin, location, and timescales of the entropy modification and of the cluster assembly process itself.

\begin{figure} [ht]
\begin{center}
\begin{tabular}{cc} 
\includegraphics[height=4cm]{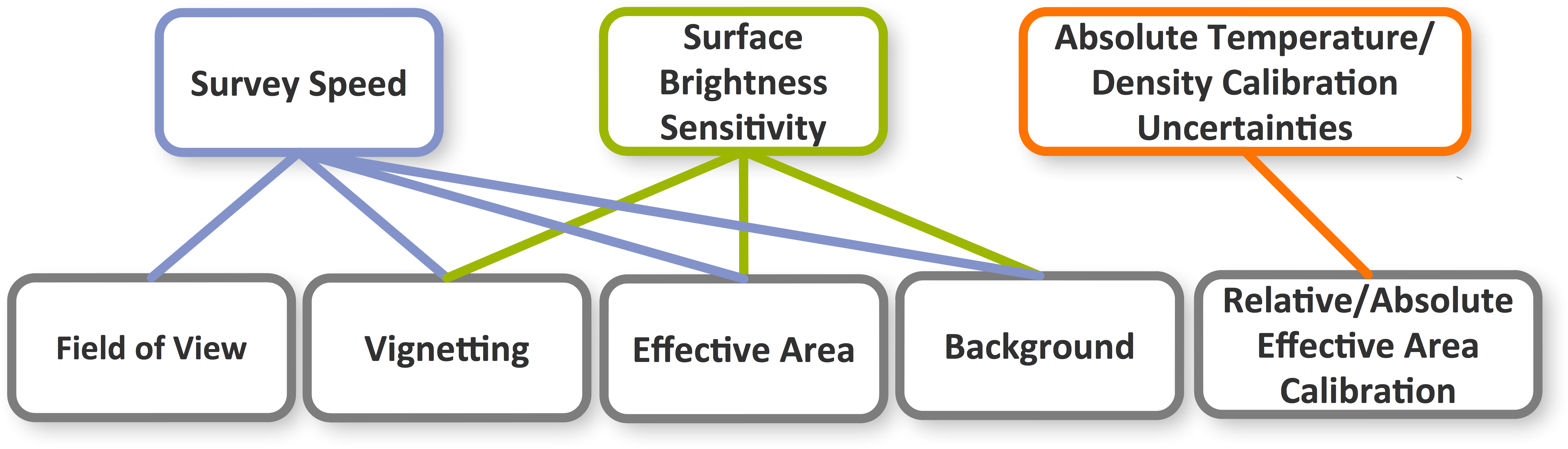}
\end{tabular}
\end{center}
\caption[example] {Relationships between the most driving requirements for the galaxy cluster outskirts science goal. The top row shows the mission independent performance requirements while the bottom row contains the derived mission/instrument specific core science requirements. Notes: surface brightness sensitivity refers here to the sensitivity over a large section of the FoV. Detailed values are presented in Table~\ref{tab:requirements}.
\label{fig:cluster_requirements} }
\end{figure}

The WFI shall provide access to spatially resolved entropy measurements at and beyond R$_{500}$ for a sample of $\sim100$ clusters of galaxies over a wide range of masses and redshift ($0<z<2$). The entropy will be obtained from spectral measurements of the gas densities and gas temperatures of extended low-surface brightness regions. This places a requirement on the surface brightness sensitivity, which predominantly depends on the effective area, vignetting and the background. Here, the background includes the contributions produced by X-ray stray light as well as the instrumental particle background. The typical angular size for high-mass, low-redshift ($z<0.2$) objects (median $R_{200}\sim24^{\prime}$) drives the need for a high survey speed, i.e., the solid angle over which a certain sensitivity is reached in a given integration time, and thus the field of view. Another important requirement is the point source sensitivity to resolve a sufficiently large fraction of the point sources forming the cosmic X-ray background and, at higher redshifts ($z>1$), to disentangle their emission from that of the extended cluster gas. Finally, very good calibrations of the absolute and relative effective areas are mandatory to achieve the necessary absolute flux, i.e., density ($<1$\,\%), and temperature ($<5$\,\%) calibration requirements. The relation between the key requirements is sketched in Fig.~\ref{fig:cluster_requirements}. The requirement values are summarised in Table~\ref{tab:requirements}.

\begin{figure} [ht]
\begin{center}
\begin{tabular}{cc} 
\includegraphics[width=16cm]{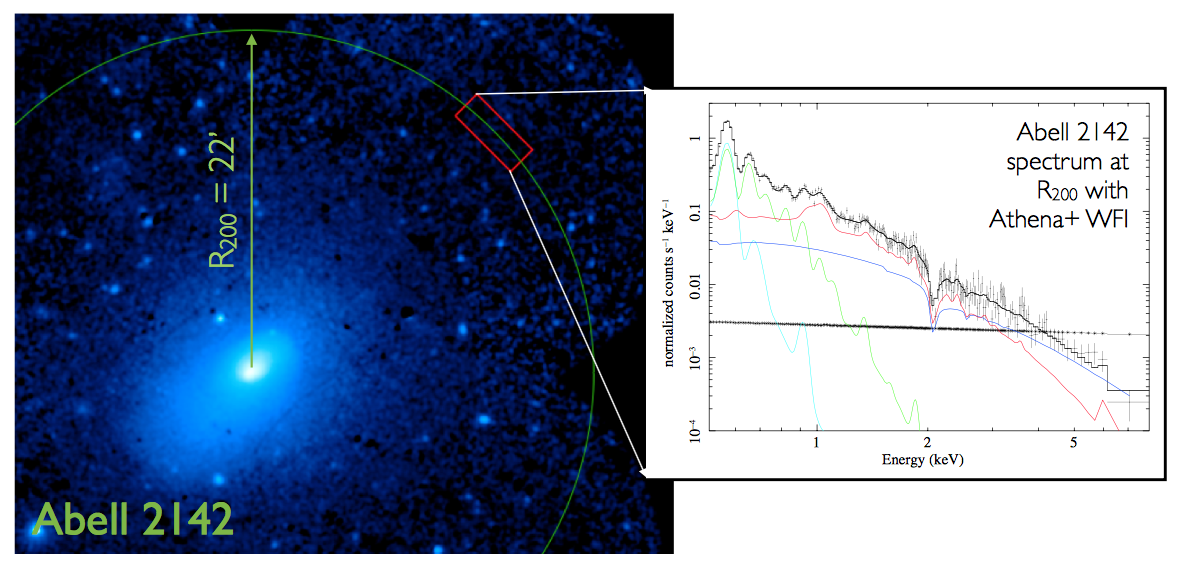}
\end{tabular}
\end{center}
\caption[example] {{\it Left:} Mosaic image of the $z=0.09$ galaxy cluster Abell~2142 taken with the {\it XMM-Newton} satellite. R$_{200}$, corresponding to a projected distance of $22^{\prime}$, is indicated by the green circle. {\it Right:} Simulated WFI spectrum of a small region at R$_{200}$ (red box in left Figure) with three components of the X-ray background (light blue, green, and dark blue lines), the instrumental background (horizontal black line), the intrinsic cluster spectrum (red line), and the total observed spectrum with the instrument background subtracted (black data points). Figures from [\citenum{Ettori:2013aa}].
\label{fig:cluster_image} }
\end{figure}

\newpage

\subsection{Spins of Stellar Mass Black Holes}
\label{subsec:spin}
 
Stellar mass black holes in Galactic binary systems can be considered as local, scaled-down versions of the supermassive black holes hosted in the centres of most galaxies (see \S~\ref{subsec:agn}). Formed as remnants of energetic stellar explosions, i.e., supernovae and gamma-ray bursts, they typically accrete matter from a non-degenerate stellar companion and occasionally even show outflows and jets with relativistic speeds. The much shorter dynamical times scales in stellar mass black holes provide access to accretion flows and jet launching processes on time scales that are otherwise inaccessible in SMBH. 

Galactic black hole binaries are among the brightest X-ray sources in the sky, thus providing ample photon flux for detailed spectral studies on short time scales. One of the main questions that can be addressed in these studies in the link between the accretion disk and the ejection of relativistic jets [e.g., \citenum{Motch:2013aa,Miller:2007aa}]. A critical parameter here is the black hole spin, whose energy could be tapped to power the outflows through magnetic connections to the ergosphere [\citenum{Blandford:1977aa}]. Measurements of the black hole spins can also be translated into improved constraints on the formation scenarios of stellar mass black holes, as it is expected that the spin received during the birth of the black hole is not significantly altered during the subsequent binary evolution.

\begin{figure} [ht]
\begin{center}
\begin{tabular}{cc} 
\includegraphics[height=8.0cm]{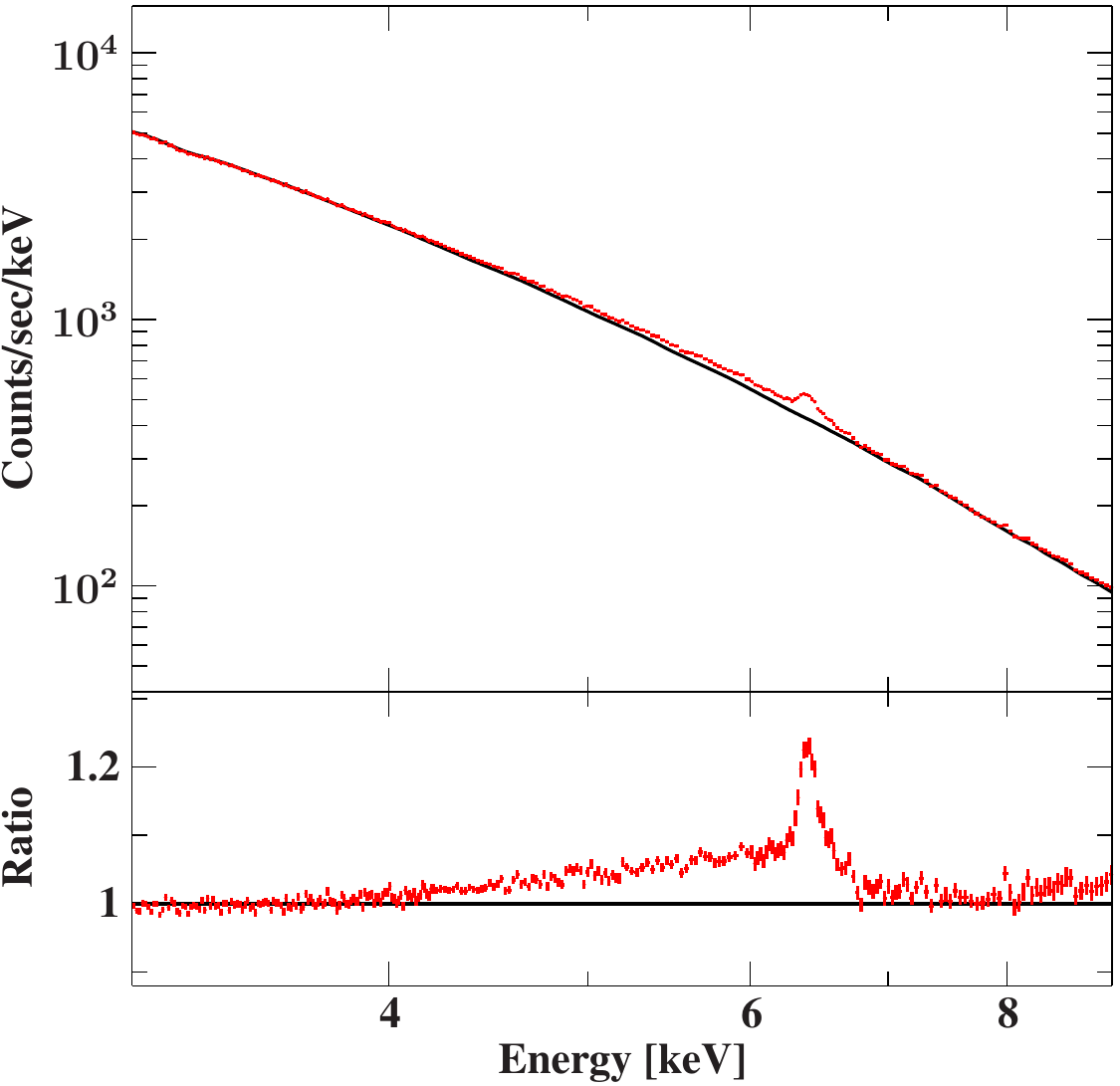}
\includegraphics[height=8.0cm]{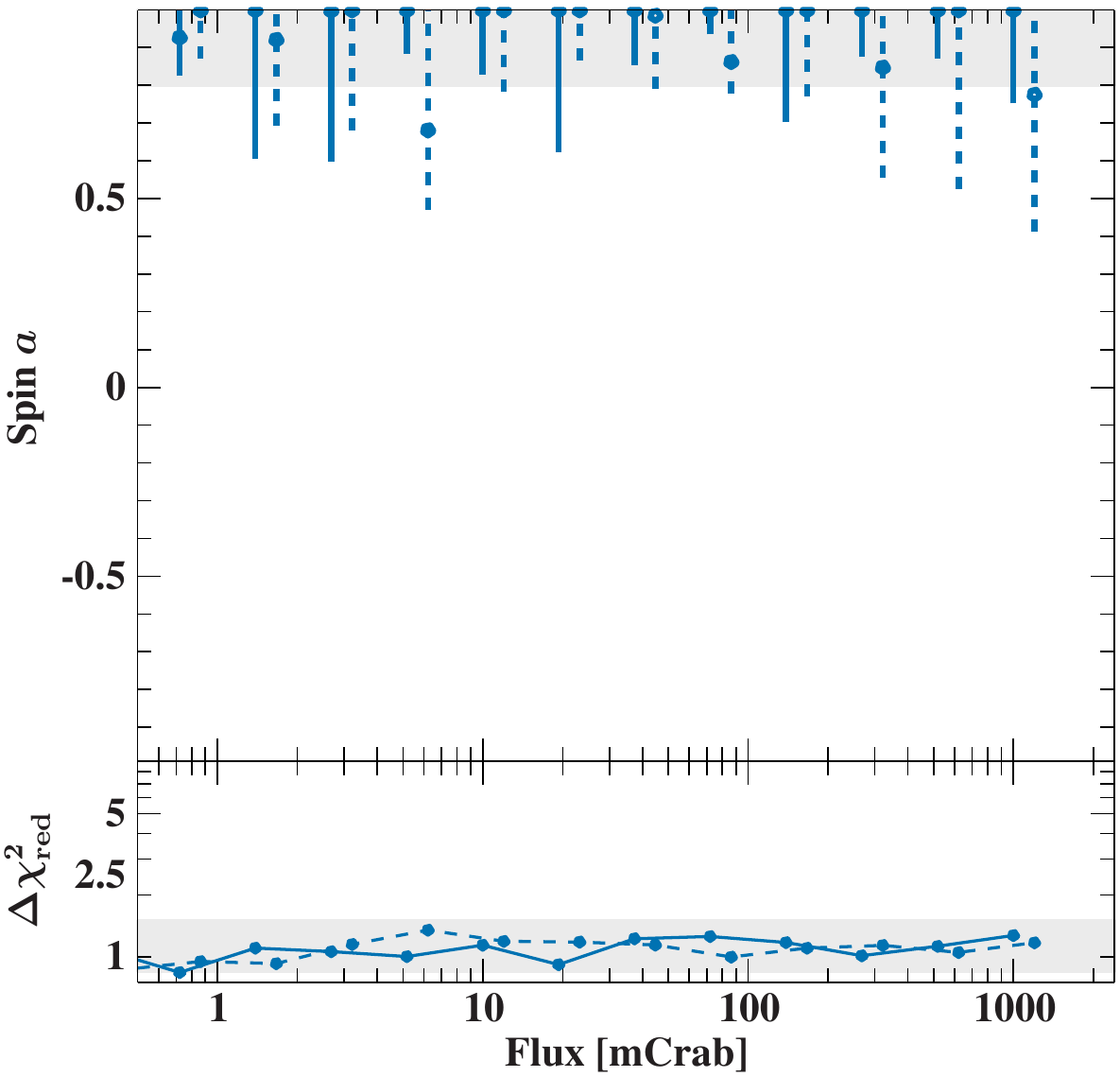}
\end{tabular}
\end{center}
\caption[example] {{\it Left:} Simulation of a 1\,ks WFI observation of a maximally co-rotating black hole (input spin $a=0.99$) black hole with a brightness of 1\,Crab and an inclination with respect to the observer of 30$^o$. Shown is the predicted count rate (red), the best fit model without Fe $K_{\alpha}$ line (red), and the ratio of data and best fit model with the residual Fe emission line complex. {\it Right:} Spin as reconstructed from spectral simulations with different source fluxes. All spectra have the same number of net counts (500,000), i.e. are integrated over different time intervals. This demonstrates the impact of detector effects independent of the signal-to-noise ratio. The solid/dashed lines show the results for two different {\it Athena} mirror models (solid: 1469mm radius, dashed: 1196mm radius), indicating the impact of the available effective area. 
\label{fig:spin_spectrum} }
\end{figure}

The black hole spin can only be accessed indirectly by actually measuring the inner radius of the accretion disk\footnote{For a non-rotating black hole, the innermost stable circular orbit (ISCO) is $R_{\rm ISCO}=6GM/c^2$ while for a co-roating maximally spinning black hole $R_{\rm ISCO}= 1.25GM/c^2$.}. This is typically achieved through X-ray observations \footnote{For the supermassive black hole in the Galactic Center and in M87, spin measurements have also been possible with sub-mm VLBI imaging. (e.g., [\citenum{Doeleman:2012aa}]). This is, however, not feasible for Galactic black hole binaries due to their significantly smaller angular size.} of the innermost stable orbit, where the accretion flow reaches the highest density and has the highest rotational velocity. Possible means are the observations of high-frequency quasi-periodic oscillations, modelling of the thermal disk emission continuum, X-ray polarisation, and/or the characterisation of atomic emission features produced by reflection of hard X-ray photons on the accretion disk.  For the latter, Iron K$_{\alpha}$ emission lines play a particularly important role due to their emission strength and favourable location in X-ray band. The line profile is complex and shaped by Newtonian line broadening,special relativity effects due to the high orbit velocities (transverse Doppler shift and beaming-enhanced blue wing) and by general relativity effects near the innermost stable orbit (extended red wing from photons losing energy in the deep black hole potential) (Fig.~\ref{fig:spin_spectrum}).
 
The WFI shall measure the spins of at least 10 Galactic black holes and 10 neutron stars. To achieve the necessary spectral quality a high photon flux, e.g., $>200000$ photons between $2-10$\,keV, is desirable [\citenum{Guainazzi:2006aa}]. While the brightest Galactic black hole binaries readily provide this required flux, observations with currently operating facilities are significantly restricted by technical limitations. The relevant metric here is the count rate capability, or more precise the product of the effective area and the time in which the detector can register incoming X-ray photons and assign the correct energies. This product is typically of the order of a few percent for existing instruments, being limited predominantly by telemetry and detector dead time. It is relevant to note, that most black hole binaries are extremely variable on short time scales and losses due to low efficiency can thus not be compensated by an increased exposure time. {\it Athena} with its large effective area and high telemetry volume together with the WFI with its dedicated fast detector will allow an $\sim100\times$ increase in the bright source capability compared to {\it Chandra} and {\it XMM-Newton}. WFI's high throughput and low pile-up  is achieved by defocusing the fast detector to spread the photon flux more homogeneously over the entire sensitive area and by reading it out with very high frame rate, i.e. with high time resolution (see \S~\ref{sec:wfi_instrument}). A good spectral resolution is required to properly measure the skewed Iron K$_{\alpha}$ line profile. The relation between the key requirements is sketched in Fig.~\ref{fig:bh_requirements}. The requirement values are summarised in Table~\ref{tab:requirements}.

\begin{figure} [ht]
\begin{center}
\begin{tabular}{cc} 
\includegraphics[height=4cm]{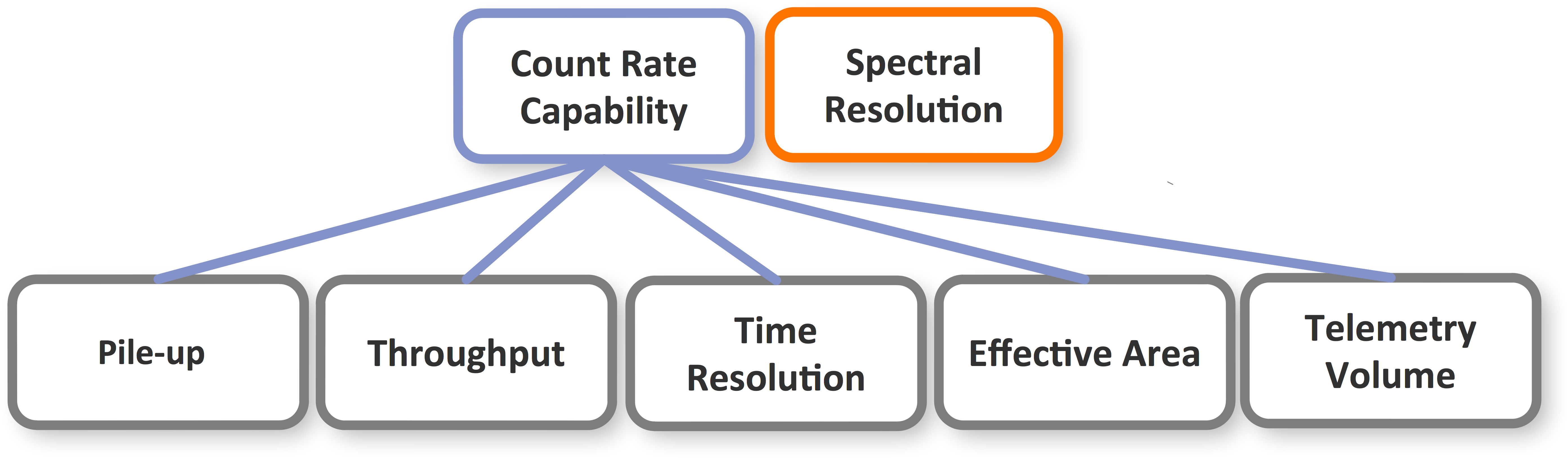}
\end{tabular}
\end{center}
\caption[example] {Relationships between the most driving requirements for the stellar mass black hole spin science goal. The top row shows the mission independent performance requirements while the bottom row contains the derived mission/instrument specific core science requirements. Note, that the spectral resolution requirement traces directly to lower level instrument requirements not shown here. Detailed values are presented in Table~\ref{tab:requirements}.
\label{fig:bh_requirements} }
\end{figure}

\clearpage

\section{Summary}
\label{sec:summary}

In this paper we briefly described three  {\it Athena}/WFI science objectives which exemplary stand for the breadth of the anticipated {\it Athena} exploration of the Hot and Energetic Universe. A full description of all the WFI science goals is for obvious reasons beyond the scope of this presentation. Using the three examples of the formation and growth of supermassive black holes (\S~\ref{subsec:agn}), non-gravitational heating in clusters of galaxies (\S~\ref{subsec:clusters}) and spin measurements of stellar mass black holes (\S~\ref{subsec:spin}), we also motivated some of the driving science requirements at mission independent and mission/instrument dependent level. A quantitative summary of those requirements is presented in Tab.~\ref{tab:requirements}.

\begin{table}[ht]
\caption{Selection of WFI key science and instrument requirements. Note, that this table only shows a subset of the requirements and that all requirements and their values are still subject to revision.}
\label{tab:requirements}
\begin{center}
\begin{tabular}{|l|l|l|}
\hline
{\bf Description} & {\bf Condition} & {\bf Value} \\
\hline
Grasp & 1\,keV & $>0.38$\,m$^2$deg$^2$\\
& 7\,keV & $>0.014$\,m$^2$deg$^2$\\
\hline
\hspace{0.5truecm}Field of View & large detector & $>40^{\prime}\times40^{\prime}$\\
\hline
\hspace{0.5truecm}Effective Area & 0.2\,keV & $>0.11$\,m$^2$\\
& 1\,keV & $>1.80$\,m$^2$\\
& 7\,keV & $>0.18$\,m$^2$\\
& 10\,keV & $>0.04$\,m$^2$\\
\hline
Survey Speed & in 100ks & $>1000$\,amin$^2$\\
\hline
Point Source Sensitivity & on-axis, 0.5-2\,keV, 450\,ks & $<2.4\times10^{-17}$\,erg cm$^{-2}$ s$^{-1}$\\
& 15$^{\prime}$ off-axis, 0.5-2\,keV, 450\,ks & $<2.4\times10^{-17}$\,erg cm$^{-2}$ s$^{-1}$\\
\hline
\hspace{0.5truecm}Point Spread Function & on-axis, 0.2-7\,keV & $<5^{\prime\prime}$ HEW\\
& 20$^{\prime}$ off-axis, 0.2-7\,keV & $<10^{\prime\prime}$ HEW\\
\hline
Surface Brightness Sensitivity & on-axis, 0.5-2\,keV, 100\,ks & $<2.4\times10^{-16}$\,erg cm$^{-2}$ s$^{-1}$ amin$^{-2}$\\
& 20$^{\prime}$ off-axis, 0.5-2\,keV, 100\,ks & $<2.4\times10^{-16}$\,erg cm$^{-2}$ s$^{-1}$ amin$^{-2}$\\
& on-axis, 5-7\,keV, 100\,ks & $<6.2\times10^{-17}$\,erg cm$^{-2}$ s$^{-1}$ amin$^{-2}$\\
& 20$^{\prime}$ off-axis, 5-7\,keV, 100\,ks & $<6.2\times10^{-17}$\,erg cm$^{-2}$ s$^{-1}$ amin$^{-2}$\\
\hline
\hspace{0.5truecm}X-ray stray light & source at 45$^{\prime}$ off-axis & suppress flux by $10^{-3}$\\
\hline
\hspace{0.5truecm}Particle Background & 1.5-7\,keV & $<5\times10^{-3}$ cnt s$^{-1}$ cm$^{-2}$ keV$^{-1}$\\
\hline
\hspace{0.5truecm}Particle Background Knowledge & 1.5-7\,keV & $<2$\,\% \\
\hline
Reconstructed Astrometric Uncertainty & large detector & $<1^{\prime\prime}$\\
\hline
Absolute Temperature Calibration & $T=5$\,keV, $z=0.5$, $Z=0.3$ solar & 4\,\%\\
\hline
Count Rate Capability & 1\,Crab & $<1$\,\%  pile-up, $>80$\,\% throughput\\
\hline
\hspace{0.5truecm}Time Resolution & fast detector & 80\,$\mu$s\\
\hline
Spectral Resolution & 1\,keV & $<80$\,eV \\
& 7\,keV & $<170$\,eV\\
\hline
\end{tabular}
\end{center}
\end{table} 

\acknowledgments 
 
The authors acknowledge the tremendous work performed by the Athena Science Study Team and the Athena Science Working Groups in defining the Athena science case and developing the science requirements. The authors are also grateful to all colleagues and institutions that supported the Wide Field Imager instrument. The work was funded by the Max-Planck-Society and the German space agency DLR (FKZ: 50 QR 1501).

\bibliography{report} 
\bibliographystyle{spiebib} 

\end{document}